# Closed-Loop Estimation of Oscillator g-Sensitivity in a GNSS/IMU System

**Vasiliy M. Tereshkov**

Topcon Positioning Systems, LLC

VTereshkov@topcon.com

**Abstract.** We propose a simple method for estimating crystal oscillator g-sensitivity in inertially aided Global Navigation Satellite System (GNSS) receivers. It does not require any specific equipment, like GNSS signal simulators or rate tables. The method is based on analyzing closed-loop phase tracking errors. This enables us to utilize the actual GNSS signal as the frequency reference, despite the presence of an unknown Doppler shift in it. The method has been successfully applied to the calibration of an oven-controlled crystal oscillator.

**I. Introduction.** Crystal oscillator $g$-sensitivity is known to have significant impact on the performance of Global Navigation Satellite System (GNSS) receivers. This topic is thoroughly studied in the recent paper [1], where it is shown that, after the $g$-sensitivity calibration, carrier phase tracking accuracy is greatly improved, particularly in the case of Phase Lock Loops (PLLs) aided by an external Inertial Measurement Unit (IMU). Nevertheless, the $g$-sensitivity measurement technique used in this research is far from easy, since it requires a very expensive GNSS signal simulator that provides a frequency reference when the oscillator is subjected to a $2g$ tip-over test.

To reduce the cost of the calibration setup, the genuine GNSS signal can be used instead of the simulated one. This, however, poses a new problem. The GNSS signal has some unknown, not necessarily constant, Doppler frequency shift due to the relative motion of the satellite and the receiver. Receiver kinematics can be eliminated by connecting it to a fixed external antenna, while the receiver mainboard with the crystal oscillator and the IMU may still be freely moved. To eliminate the effects of satellite kinematics, we propose a novel closed-loop estimation scheme, in which the crystal oscillator drives receiver PLLs and the residual carrier phase tracking errors are analyzed together with IMU measurements to obtain an estimate of the oscillator $g$-sensitivity.

**II. PLL structure.** A typical PLL is shown in Fig. 1. It consists of a Numerically Controlled Oscillator (NCO) that generates a local replica of the carrier, a discriminator that compares this replica with the actual carrier and extracts the phase tracking error, $\varphi$, and a filter that forms the desired frequency for the NCO, using $\varphi$ as the input. Any oscillator imperfections corrupt this desired frequency by adding some frequency bias, $\Delta f$, which possibly depends on applied accelerations. Fig. 1 shows, as an example, a second order PLL with the natural frequency $\omega_0$ and the damping factor $\zeta$. It is capable of tracking the carrier with a constant Doppler frequency shift perfectly, i.e., with $\varphi = 0$. If the GNSS satellite acceleration along the line-of-sight is negligible, this PLL is sufficient; otherwise, a third order PLL is used, which can also perfectly track the Doppler shift rate of change, i.e., the satellite acceleration [2].

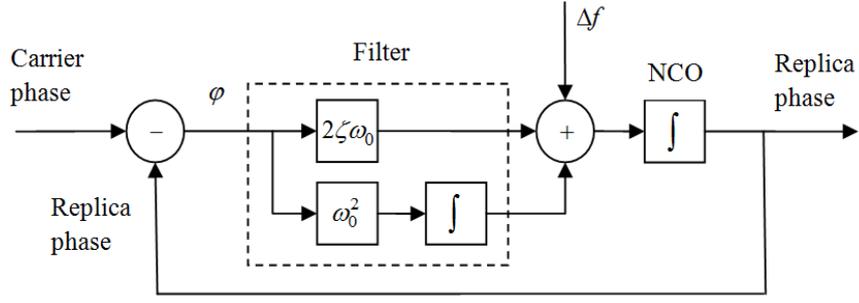

Fig. 1. PLL Block Diagram

**III. Estimator design.** In terms of phase and frequency errors, the PLL operation is essentially linear. This allows us to introduce a linear operator (transfer function) $L$ such that $\varphi = L[\Delta f]$. Its explicit form depends on the PLL order. To characterize crystal oscillator $g$-sensitivity, we apply the same Filler model as used in [1], where the frequency error is represented by the scalar product of the specific force vector $\mathbf{a}$ and the sensitivity vector $\mathbf{\Gamma}$, i.e., $\Delta f = \mathbf{a}^T \mathbf{\Gamma}$. In our study it is convenient not to normalize $\mathbf{\Gamma}$ by the nominal oscillator frequency, since we are concerned with the absolute frequency errors. When the frequency error propagates through the PLL, it yields a phase tracking error $\varphi = L[\mathbf{a}^T \mathbf{\Gamma}]$. Our purpose is to estimate $\mathbf{\Gamma}$ by measuring $\mathbf{a}$ and $\varphi$.

Consider a set of three software PLL models, $x$, $y$ and $z$. Each of them implements the same operator $L$, but is fed directly by the $x$, $y$ or $z$ component of the specific force, respectively. In other words, the sensitivity vectors for these PLL models are unit vectors that form the coordinate basis. Their output phase tracking errors constitute a vector which we denote by $\boldsymbol{\varphi}_1 = L[\mathbf{a}]$. Due to the linear properties of $L$, the actual phase tracking error can therefore be expressed as $\varphi = \boldsymbol{\varphi}_1^T \mathbf{\Gamma}$. Similarly, if we already have some estimate $\hat{\mathbf{\Gamma}}$ of the sensitivity vector $\mathbf{\Gamma}$, then the corresponding phase tracking error is $\hat{\varphi} = \boldsymbol{\varphi}_1^T \hat{\mathbf{\Gamma}}$.

The principle of $g$-sensitivity estimation is to adjust $\hat{\mathbf{\Gamma}}$ unless the phase residual $\varphi - \hat{\varphi}$ vanishes (Fig. 2). To achieve this, we propose a gradient observer

$$\dot{\hat{\mathbf{\Gamma}}} = k \boldsymbol{\varphi}_1 (\varphi - \hat{\varphi})$$

with a positive gain $k$, which can be either constant or varying according to some rule (e.g., a Kalman-type gain).

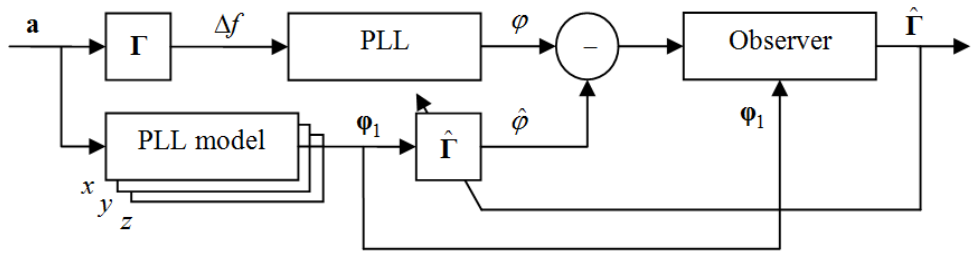

Fig. 2. Estimation Block Diagram

**IV. Estimator convergence.** If we introduce the estimation error $\tilde{\mathbf{\Gamma}} = \mathbf{\Gamma} - \hat{\mathbf{\Gamma}}$, then the error dynamics will be $\dot{\tilde{\mathbf{\Gamma}}} = -\dot{\hat{\mathbf{\Gamma}}} = -k\boldsymbol{\varphi}_1(\varphi - \hat{\varphi}) = -k\boldsymbol{\varphi}_1(\boldsymbol{\varphi}_1^T \mathbf{\Gamma} - \boldsymbol{\varphi}_1^T \hat{\mathbf{\Gamma}}) = -k\boldsymbol{\varphi}_1 \boldsymbol{\varphi}_1^T \tilde{\mathbf{\Gamma}}$. Consider a positive-definite Lyapunov function candidate $V = (1/2)|\tilde{\mathbf{\Gamma}}|^2$. Its time derivative is $\dot{V} = \tilde{\mathbf{\Gamma}}^T \dot{\tilde{\mathbf{\Gamma}}} = -k\tilde{\mathbf{\Gamma}}^T \boldsymbol{\varphi}_1 \boldsymbol{\varphi}_1^T \tilde{\mathbf{\Gamma}} = -k\left(\boldsymbol{\varphi}_1^T \tilde{\mathbf{\Gamma}}\right)^2$, provided that $k$ is constant. The scalar product $\boldsymbol{\varphi}_1^T \tilde{\mathbf{\Gamma}}$ can be treated as the projection of $\tilde{\mathbf{\Gamma}}$ to the direction of $\boldsymbol{\varphi}_1$. Suppose that the $x$, $y$ and $z$ components of $\boldsymbol{\varphi}_1$ are changing independently. Then this projection reaches zero only when $\tilde{\mathbf{\Gamma}}$ is also zero. Otherwise, $\dot{V} < 0$, so the equilibrium state $\tilde{\mathbf{\Gamma}} = \mathbf{0}$ is stable.

**V. Practical considerations.** By construction, a PLL is insensitive to any constant frequency bias $\Delta f$ (Section II). On the one hand, this property enables us to utilize the GNSS signal with some unknown Doppler shift as a reference for oscillator $g$-sensitivity estimation. On the other hand, it prevents the use of a simple $2g$ tip-over test, since in both positions ($+g$ and $-g$) the steady-state phase tracking error is zero and the sensitivity vector is unobservable. To achieve observability, one should apply variable specific force in some nearly periodic motion – either rotational or oscillating – in three orthogonal directions independently. There is no need to have perfect periodicity or an exactly known magnitude of the specific force, so this can be easily done manually, without requiring any specific equipment like rate tables. Acceptable frequency range of specific force oscillations is however limited by the PLL frequency response, $\Delta f$ being the input and $\varphi$ the output of the PLL. Its example is shown in Fig. 3 for some particular PLL parameter values: $\zeta = 1/\sqrt{2}$, $\omega_0/(2\pi) = 0,9$ $Hz$, which correspond to noise bandwidth $B_n = 3$ $Hz$. It is clearly seen that the best observability is achieved when the applied specific force is changing at the natural frequency of the PLL.

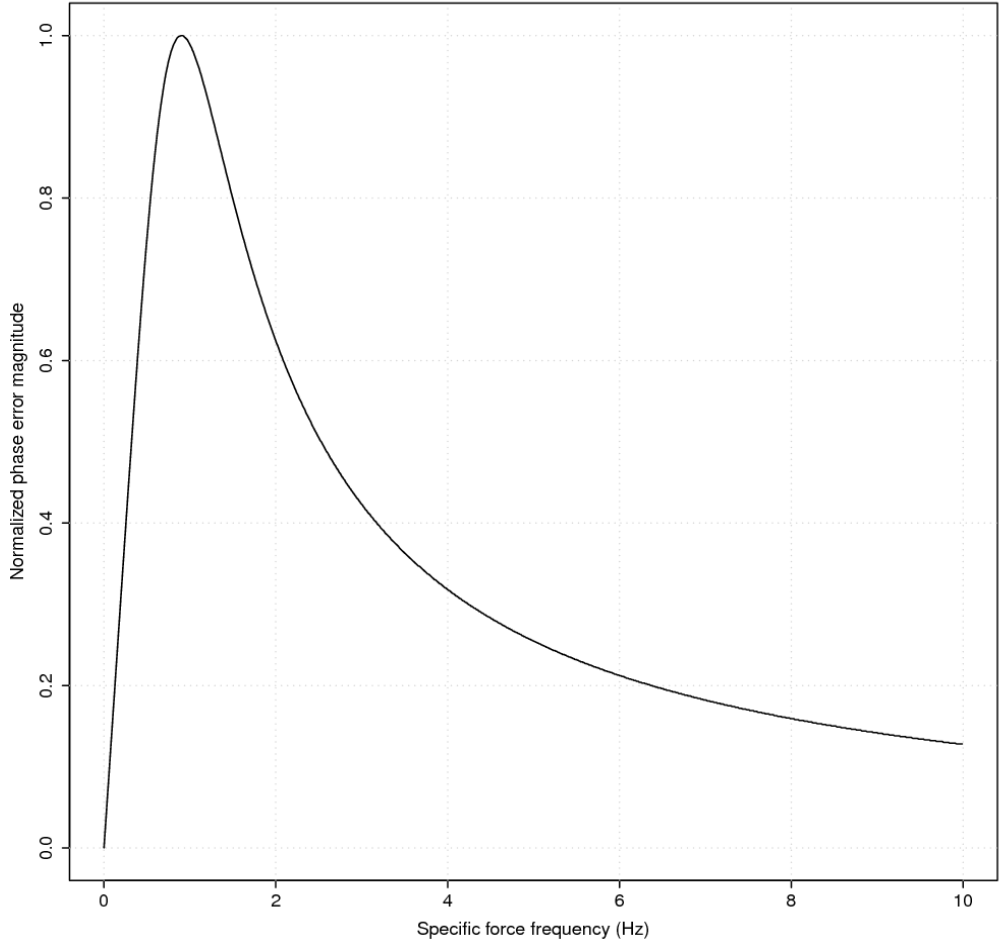

Fig. 3. PLL frequency response

The PLL noise bandwidth during $g$-sensitivity calibration should generally be the same as in the conditions for which the GNSS receiver is designed. When the bandwidth is high ($B_n > 20\ Hz$), which is typical for unaided PLLs in dynamic scenarios, the effects of $g$-sensitivity are negligible and therefore need not be estimated. An extremely low bandwidth ($B_n < 2\ Hz$), though attractive for IMU-aided GNSS receivers, may lead to the loss of lock due to the oscillator short-term instability. Such a setting seems to be acceptable for modern Oven-Controlled Crystal Oscillators (OCXOs) but not for much cheaper and less stable Temperature Compensated Crystal Oscillators (TCXOs). The above-mentioned value $B_n = 3\ Hz$ is proposed as a trade-off suitable for both OCXOs and TCXOs.

Up to now we have been studying the estimation scheme based on a single PLL. A GNSS receiver, however, runs multiple PLLs simultaneously, at least one per each satellite tracked. All these PLLs are driven by the same crystal oscillator, so the carrier phase tracking errors caused by oscillator imperfections are also the same, provided that the PLLs have equal bandwidths. Other error sources result in phase noise without any cross-correlation between PLLs. It is therefore beneficial to perform phase error averaging over several PLLs, which are selected, for example, by satellite elevation or by signal strength. The averaged phase tracking error can then serve as the input of the estimator.

**VI. Experimental validation.** The proposed estimator was used for $g$-sensitivity calibration of a Morion OCXO installed in a Topcon AGI-4 GNSS/IMU integrated receiver. The receiver was connected to a fixed external antenna and was manually moved in an oscillating manner in three orthogonal directions with a period of about $1...2\ s$. Receiver PLL noise bandwidth was set to $3\ Hz$. Phase tracking errors for all active PLLs were logged at $40\ Hz$ synchronously with specific force measurements provided by the IMU. These logs were post-processed to estimate oscillator $g$-sensitivity. Four Global Positioning System (GPS) satellites with highest L1 signal-to-noise ratios were selected, and the corresponding phase tracking errors were averaged with equal weights before being input to the estimator. The calibration results are presented in Fig. 4. The final sensitivity vector estimate is $\hat{\mathbf{\Gamma}} = [+0.09,\ +0.08,\ -0.08]^T\ Hz/(m/s^2)$.

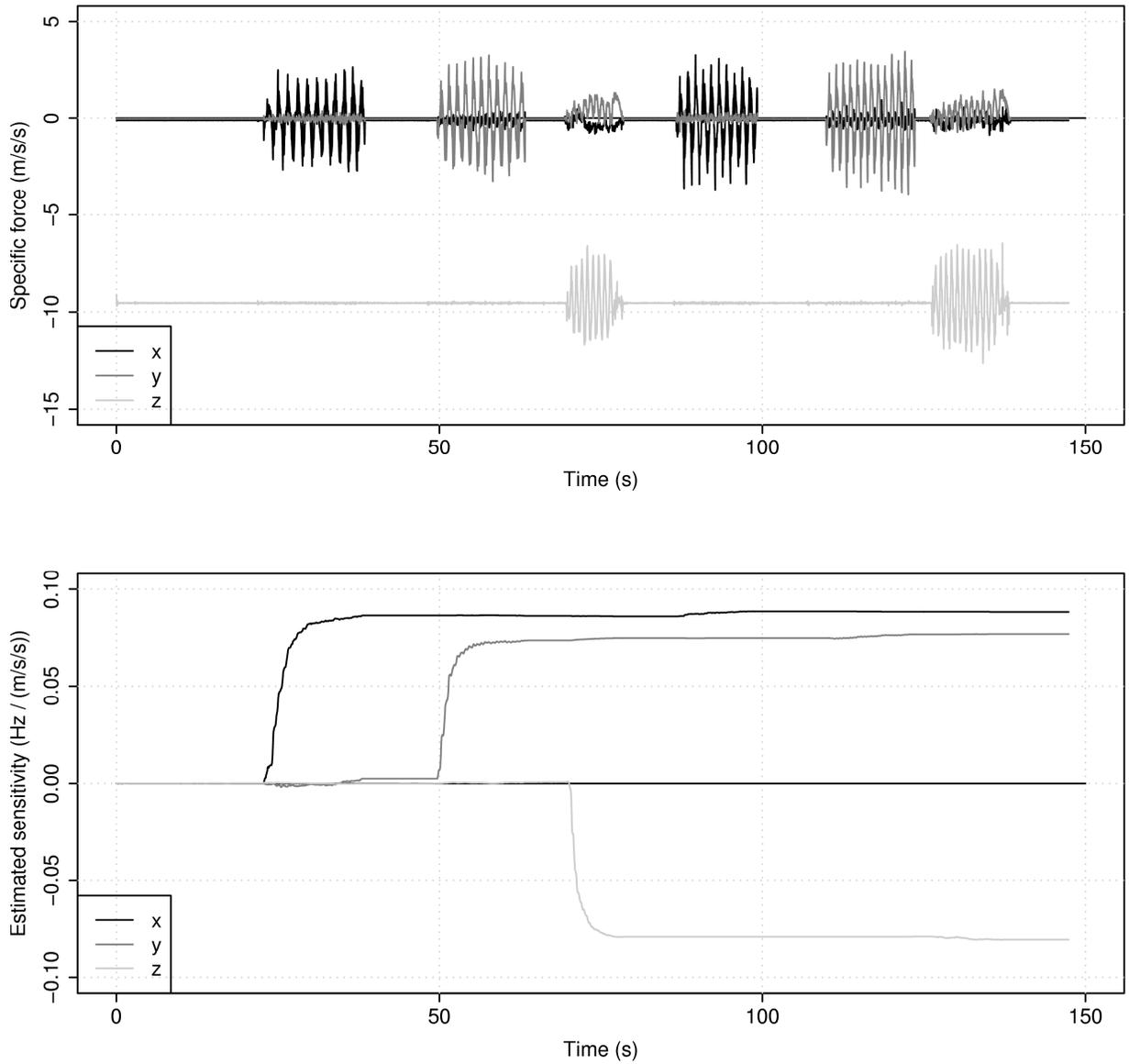

Fig. 4. Estimation process: specific force components (top), estimated sensitivity vector components (bottom)

From a practical standpoint, oscillator $g$-sensitivity manifests itself in carrier phase tracking errors. Hence, the best way to assess $g$-sensitivity estimator performance is to compare phase tracking errors before and after $g$-sensitivity compensation, using the actual GNSS signal as a reference. This comparison is shown in Fig. 5. When the GNSS receiver is at rest, $g$-sensitivity compensation has no effect, and the phase error – caused primarily by the oscillator instability over time – is the same in both cases. When the receiver is subjected to a variable acceleration, the uncompensated PLL exhibits a dramatic growth of phase error, whereas the compensated PLL does not. This confirms the correctness of the proposed estimation method.

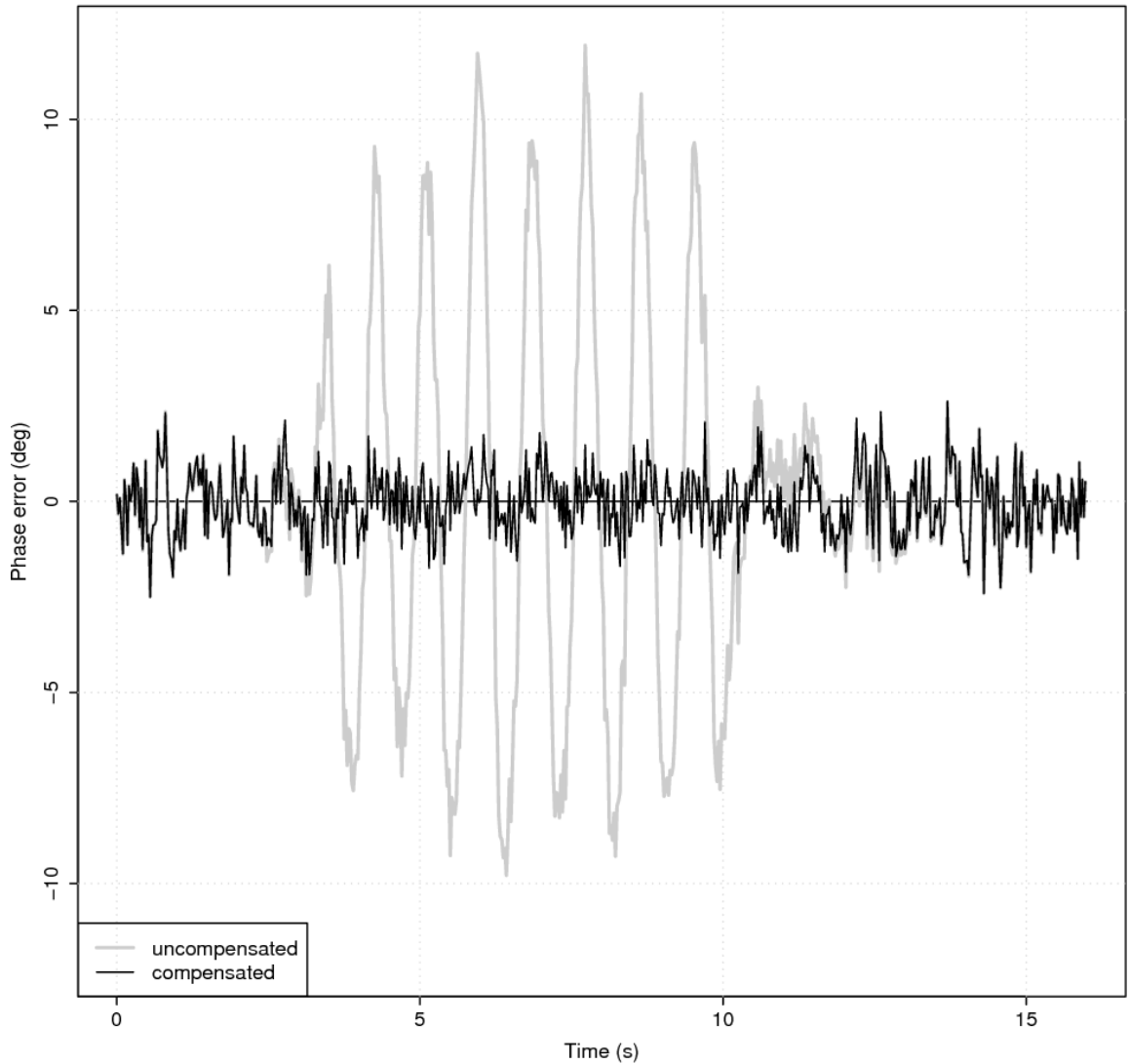

Fig. 5. The effect of *g*-sensitivity compensation

**VII. Conclusions.** In GNSS receivers, crystal oscillator *g*-sensitivity can be estimated without requiring any specific hardware such as GNSS signal simulators or rate tables. To perform this, we proposed a closed-loop estimation method. In this scheme the oscillator drives receiver PLLs, and their averaged phase tracking error is compared with the output of a triad of software PLL models fed by the same specific force components as the real PLLs, but with an adjustable oscillator *g*-sensitivity. This *g*-sensitivity is tuned unless the modeled phase error response coincides with the measured one. The proposed estimator design was verified both by theory and experiments with a Topcon AGI-4 GNSS/IMU integrated receiver equipped with a Morion OCXO.